\documentclass[preprint]{aastex631}
\accepted{for publication in \apj}

\shorttitle{Where are the Water Worlds?}
\shortauthors{Chakrabarty \& Mulders}
\usepackage{natbib}
\usepackage{calrsfs}
\usepackage{graphicx}
\graphicspath{{figures/}}

\begin{document}

\newcommand{\re}{$R_\oplus$}
\newcommand{\me}{$M_\oplus$}
\newcommand{\de}{$\rho_\oplus$}
\newcommand{\unif}{\mathcal{U}}
\newcommand{\f}{\mathcal{F}}
\newcommand{\ho}{H$_{\rm 2}$O }
 
\title{Where are the Water Worlds? Identifying the Exo-water-worlds Using Models of Planet Formation and Atmospheric Evolution}

\correspondingauthor{Aritra Chakrabarty}
\email{aritra.chakrabarty@dataobservatory.net}

\author[0000-0001-6703-0798]{Aritra Chakrabarty}
\affiliation{Data Observatory Foundation, DO;  Diagonal Las Torres N$^\circ$2640, Building E, Pe\~{n}alol\'{e}n, Santiago, Chile}
\affil{Facultad de Ingenier\'ia y Ciencias, Universidad Adolfo Ib\'a\~nez, Av.\ Diagonal las Torres 2640, Pe\~nalol\'en, Santiago, Chile}
\affiliation{Millennium Institute for Astrophysics, Chile}

\author[0000-0002-1078-9493]{Gijs D. Mulders}
\affil{Facultad de Ingenier\'ia y Ciencias, Universidad Adolfo Ib\'a\~nez, Av.\ Diagonal las Torres 2640, Pe\~nalol\'en, Santiago, Chile}
\affil{Millennium Institute for Astrophysics, Chile}

\begin{abstract}
Planet formation models suggest that the small exoplanets that migrate from beyond the snowline of the protoplanetary disk likely contain water-ice-rich cores ($\sim 50\%$ by mass), also known as the water worlds. While the observed radius valley of the Kepler planets is well explained with the atmospheric dichotomy of the rocky planets, precise measurements of mass and radius of the transiting planets hint at the existence of these water worlds. However, observations cannot confirm the core compositions of those planets owing to the degeneracy between the density of a bare water-ice-rich planet and the bulk density of a rocky planet with a thin atmosphere. We combine different formation models from the Genesis library with atmospheric escape models, such as photo-evaporation and impact stripping, to simulate planetary systems consistent with the observed radius valley. We then explore the possibility of water worlds being present in the currently observed sample by comparing them with the simulated planets in the mass-radius-orbital period space. We find that the migration models suggest $\gtrsim 10\%$ and $\gtrsim 20\%$ of the bare planets, i.e. planets without primordial H/He atmospheres, to be water-ice-rich around G- and M-type host stars respectively, consistent with the mass-radius distributions of the observed planets. However, most of the water worlds are predicted to be outside a period of 10 days. A unique identification of water worlds through radial velocity and transmission spectroscopy is likely to be more successful when targeting such planets with longer orbital periods.
\end{abstract}

\section{Introduction} \label{sec:intro}
Kepler observations have shown that the small exoplanets with radius between 1 \re~ and 4 \re~ are extremely common among the planets with orbital periods shorter than 100 days \citep[e.g.,][]{lissauer11b, mulders15b, mulders+18, he21}. Mass and radius measurements of such planets show evidence of two populations of planets, rocky and gaseous planets \citep{lopez14, rogers15}, from the break in the mass-radius relation \citep[e.g.,][]{wolfgang16, chen17}. Futher studies with improved precision in stellar radii from the follow-up surveys like the California-Kepler survey (CKS) and with the help of Gaia improved parallaxes \citep{johnson17, vaneylen18} showed that the size distribution of the small exoplanets is bimodal with a radius gap, also known as the ``radius valley", at $\sim$1.8-2 \re~ \citep{fulton17, fulton18, ho23}. This suggests a clear bifurcation between the ``super-Earth" population at the lower peak and the ``sub-Neptune" population at the higher peak. 

The two leading theories that have emerged to explain these two populations of exoplanets are atmospheric dichotomy and compositional dichotomy. The first theory suggests that sub-Neptunes are the planets with primordial H/He dominated atmospheres and super-Earths are the planets that have lost their primordial atmospheres. Mass-loss mechanisms that can cause small planets to lose their primordial atmospheres entirely include photo-evaporation loss due to high energy stellar flux \citep[e.g.,][]{owen17, owen20, rogers21, rogers23}, core-powered mass loss \citep[e.g.,][]{ginzburg18, gupta19}, giant impact loss \citep[e.g.,][]{inamdar16, biersteker19, chance22}, among others. The other theory suggests that the size bimodality is a manifestation of the two types of core compositions of the small planets. Super-Earths are the planets with Earth-like rocky core composition and sub-Neptunes are the planets with water-ice-rich cores with larger radii due to lower density \citep[e.g.,][]{mordasini09, raymond18a, zeng21, venturini20}. These water-ice-rich planets, also known as ``water worlds", are the planets that form exterior to the snowline of the protoplanetary disks and migrate inward through type-I migration to the location where we observe today \citep[e.g.,][]{izidoro17, raymond18a, raymond18b}. Such plantes form with a silicate-to-ice ratio of 1:1 \citep[e.g.,][]{lodders03, lopez17, zeng19, aguichine21,  mousis20} beyond the snowline. These are analogous to the water-rich minor planets and moons (e.g., Enceladus, Pluto, etc.) of the solar system and the water-rich cores of Uranus and Neptune \citep{mousis18}. However, the existence of such planets around other stars at short orbital distances remains elusive to date.

\cite{luque22} presented a sample list of small (size $<$ 4 \re) transiting planets around M-dwarfs with refined mass and size estimation which strongly hints at a distinct population of these water worlds in the mass-radius and density space. However, the observed mass-radius distributions can also be explained with a rocky population of planets formed in-situ with varying atmospheric content \citep{rogers23}. Moreover, in-situ models have been successful at explaining the broad distributions of orbital periods, eccentricities, and radii of the Kepler planets \citep[e.g.,][]{hansen13, ogihara15, macdonald20}. So the presence of water worlds is not a foregone conclusion from current observations.

However, in-situ planets start with excess mass in the inner region of the disks, and therefore need either pebble drift \citep[e.g.,][]{boley14, chatterjee14, chatterjee15} or planet migration \citep[e.g.,][]{izidoro17, raymond18a, raymond18b}, which would also bring in water worlds from beyond the snowline to the inner region. The latter processes also include planet-disk gravitational interaction that in-situ models ignore \citep{izidoro17}. Moreover, migration models too can explain most features of the Kepler systems, including the radii and mutual inclinations (coplanarity) of the planets, and the period-ratio distribution of the systems in mean-motion resonance (e.g., Trappist-1 system). However, most (90-95\%) Kepler systems are found to be somewhat offset from the first-order mean-motion resonance \citep{lissauer11a, fabrycky14} which would require other physical processes, such as dynamical instabilities and collisions \citep{izidoro17, izidoro22} among others, to break the resonance chains after the migration phase.

While the planets containing rocky cores have been shown to reproduce the Kepler size bimodality with the help of atmospheric escape models, the possible fate of the hypothetical water worlds after migrating inward has been somewhat contentious. Many of the early models on compositional dichotomy do not provide any explanation to why the water worlds would not accrete H/He dominated atmospheres \citep{izidoro17, raymond18a}. \cite{izidoro22} present their pebble accretion model of planet formation to show that giant impacts on the migrating planets by planetesimals can wipe away the primordial atmospheres of most of the planets, unveiling the two kinds of core compositions for a certain initial condition. In such a case, giant impacts are shown to dominate over other atmospheric mass-loss mechanisms, such as photo-evaporation, responsible for atmospheric dichotomy. Although both photo-evaporation and giant impacts can account for the atmospheric erosion of super-Earths, measuring the D/H ratio of individual planets' atmospheres presents a potential method to differentiate between the two escape methods. The fractional escape of the primordial H/He envelope in the case of photo-evaporation is known to alter the atmospheric D/H ratio \citep{gu23}. In contrast, impact-driven escape may result in either no fractionation or a distinct level of fractionation. On the other hand, \cite{mordasini18} discusses the presence of water worlds having atmospheres by using the Bern model but most of such planets turn out to be Neptune-sized and hence, not useful in reproducing the Kepler size bimodality. Hence, the impact of the atmospheric evolution and escape of the water-rich sub-Neptunes with atmospheres on the overall demographics of small exoplanets remains a subject of interest. This is further motivated by the recent developments in the model of internal structures of water worlds with H/He atmospheres \citep{lopez17, zeng19}.

In this paper, we provide an agnostic overview of the effect of atmospheric evolution and mass-loss processes such as photo-evaporation and giant impacts on the planets simulated through two suites of formation models: migration and in-situ. We use the planetesimal accretion models from the Genesis library developed by \cite{mulders20} for that purpose. We compare the outcomes of our models with both the size distribution from Kepler and mass-radius distributions of the transiting planets that have been followed up by radial velocity (RV) observations (e.g., LP22). We further use these comparisons to benchmark our migration models from which we predict the occurrences of water worlds as a function of their mass, radius and orbital period. In Section~\ref{sec:meth}, we explain the general methodology: the combinations of formation models, host stars, and atmospheric mass-loss processes and how we implement them to reproduce the Kepler size bimodality. We define the possible water worlds from observational perspective and discuss the occurrences of such water worlds from observations (as upper limits) and from models in Section~\ref{sec:ww}. In Section~\ref{sec:guide}, we elaborate on how we calculate the probabilities of finding water worlds from their kernel density estimations and use them to guide future search for water worlds. We conclude the key points in Section~\ref{sec:con}.

\section{Methods}\label{sec:meth}
Here we study the different processes of atmospheric evolution of the small planets in tandem with their formation theories. Instead of adopting some initial random distributions for the properties of the planetary cores in our study, we use the outcomes of the planetesimal accretion models available from the Genesis database (see Section~\ref{sec:meth/gen}). This combined study provides an avenue to analyze the possible bulk compositions of the observed planets from their bulk density and orbital period distributions. This study also allows us to verify which of the contesting theories of atmospheric evolution of the small planets effectively reproducing the Kepler size bimodality are consistent with the formation theories. Moreover, the Genesis models include migrating planets in addition to planets forming in-situ and also simulate the giant impact phases required to study the effect of impact loss.
We produce different models for the final architectures of the simulated planets based on their formation process (migration or in-situ), type of the host star (G or M), atmospheric loss mechanisms (photo-evaporation or impact), and bulk composition (rocky or water-ice-rich). 

\subsection{The Genesis Database of Models of Planet Formation}\label{sec:meth/gen}

The Genesis database\footnote{\url{https://eos-nexus.org/genesis-database/}} is a library of N-body simulations of formation of the terrestrial planets developed by \cite{mulders20}. These models use the simulations of \cite{hansen13} to model the Kepler super-Earths as a starting point. Based on the distribution pattern of the solids around the host stars, the annular range of distribution of the solids, and the ratio of the number of embryos to the number of planetesimals, \cite{mulders20} presented 15 models with different initial conditions, each having 50 runs of simulations. 12 of such models include planets growing in-situ and the other 3 include migration. Each model shows a different distribution of core mass and orbital period (see Figure 3 of \citealt{mulders20}). We combine these models so that the final suites of models exhibit lognormal-like mass distributions ranging between $\sim1$\me~ and $\sim25$\me~ and period distributions closest to the Kepler period distribution ranging between $\sim1$ day and $\sim100$ days. We arrive at 3 suites of such models:
\begin{itemize}
\item In-situ: combining all 12 in-situ models from Genesis.
\item Migration-A: combining the migration models Gen-M-s22 and Gen-M-s50 from Genesis.
\item Migration-B: combining the migration models Gen-M-s10 and Gen-M-s50 from Genesis.
\end{itemize}
All three migration models in Genesis, namely Gen-M-s10, Gen-M-s22, and Gen-M-s50, initiate with solids distributed between 0.05 AU and 10 AU, albeit with different normalizations of the surface density.  They migrate inward under the influence of the gas disk that stretch upto an inner edge of 0.05 AU. Depending on the snowline location (see Section~\ref{sec:meth/host/sl}), migrating planets that make up the population of small exoplanets in the inner region of the disk are likely to consist of both rocky and water-ice-rich cores, further elaborated in Section~\ref{sec:meth/coreatm}. The gas disk disperses after 5 Myr from the start of the simulations after which dynamic instabilities set in. This instability phase leads to breaking of several of the resonant chains of planets that formed during migration, orbital crossing among planets, and giant impacts. Unlike the in-situ suite of models, the migration simulations do not incorporate lower-mass planetesimals, which could potentially impact the atmospheric erosion process, as suggested by \cite{chen22}. However, as our current study does not consider the influence of planetesimals on atmospheric erosion, their absence does not significantly affect the evolution of migrated planets. The potential impact of planetesimals is further discussed in Section~\ref{sec:meth/discuss}.

The N-body simulations in the Genesis database assume perfect accretion of the cores. This is corroborated by independent studies carried out by \cite{chambers13} and \cite{dugaro20}, where they  conducted comparisons to assess the impact of hit-and-run collisions and collisional fragmentation relative to perfect mergers. \cite{chambers13} argued that large bodies often reaccrete mantle fragments at a subsequent time, leaving their final composition largely unchanged. Conversely, for water worlds, \cite{dugaro20} found that the fragments’ contribution to their final mass and water content is negligible. The distributions of the core masses for the 3 suites of models are shown in Figure~\ref{fig:core-dist}. The observed period ratio distribution of small Kepler exoplanets (size $< 4$ \re) suggests that majority of the resonant chains break after the gas disk dispersal \citep[e.g., ][]{izidoro22, izidoro21, izidoro17}. To match our migration models with this feature, we choose 95\% of the runs with broken resonant chain and 5\% of the stable runs for each model, following \cite{izidoro21, izidoro22}.

This step only provides us with the mass and orbital period distributions of the planetary cores. To determine the bulk compositions of the cores we need to define the location of the snowline which in turn depends on the spectral type of the central host star. The choice of host stars is explained in the following subsection.

\begin{deluxetable}{cccc}[!h]
\tablecaption{Fiducial values of the properties of the host stars and the disks chosen in our models.}
\label{tab:host}
\tablehead{
\colhead{Host star} & \colhead{Mass (M$_\oplus$)} & \colhead{T$_{\rm eq}$ at 1 AU (K)} & \colhead{Snowline location (AU)}
}
\startdata
G & 1.0 & 279 & 2.2 \\
M & 0.35 & 108 & 0.8
\enddata
\end{deluxetable}

\subsection{Host Stars}\label{sec:meth/host}
The final architectures of the planets depend on the central star. We consider two types of hosts stars for our models: G-type and M-type. We use the same distribution of planetary cores adopted from the Genesis database, albeit simulated for a Sun-like star, for both the choices of the host stars. This is motivated by the fact that the overall size distribution of Kepler planets \citep{gupta19, ho23} as well as the total mass budget of the super-Earths and sub-Neptunes (see discussions in \citealt{mulders18}) does not alter significantly with the change in the host star mass, although the disk masses are known to scale with the stellar mass \citep{pascucci16}. The parameters that change with the host stars are the location of snowline of the disks which dictates the water-ice content of the cores, the luminosity of the star which dictates the equilibrium temperature ($T_{eq}$) of the planets, and the high energy XUV flux from the stars which dictate the photo-evaporation loss rate of the atmospheres of the planets. The fiducial values of the parameters for the two types of stars chosen in our models are shown in Table-\ref{tab:host}. For the G-type host star, we have chosen the same values as that of the Sun and for the M-type host star, we have chosen the median values of the parameters of the host stars in the target list of \cite{luque22}.

\subsubsection{Location of Snowline} \label{sec:meth/host/sl}

We consider the location of the snowline at the time of planetesimal formation as this is the time when the water/ice gets trapped into the planetesimals beyond the snowline. They do not further sublimate on migrating inward unlike the pebbles \citep{mulders15} but can only lose water partially by late collisions after the gas-disk dispersal or from $^{\rm 26}$Al heating \citep{izidoro17, izidoro22, raymond18a, monteux18}. The location of the snowline in the solar system at the time of planetesimal formation has been inferred from the transition between hydrous and anhydrous asteroids to be $\sim2.5$ AU \citep{abe00}. However, this estimated location of snowline could change significantly if we consider that the asteroids have been scattered into the current location \citep{walsh12, demeo14}. Moreover, \cite{mulders15} calculate the $1\sigma$ range of locations of the snowlines for a solar-mass star to be between $\sim 2$ AU and $\sim 5$ AU for a disk made up of small grains and between $\sim 1.1$ AU and $\sim 2.4$ AU for a disk made up of large grains. Keeping this uncertainty in mind, we choose the fiducial location of the snowlines for the G-type stars at 2.2 AU. Following the scaling relation of \cite{mulders15} and keeping in mind the associated uncertainties, we select the fiducial location of the snowlines for M-type stars as 0.8 AU. However, the effect of other values of snowline locations is described in Section~\ref{sec:ww/occ}.

\begin{figure}
\centering
\includegraphics[scale=0.4]{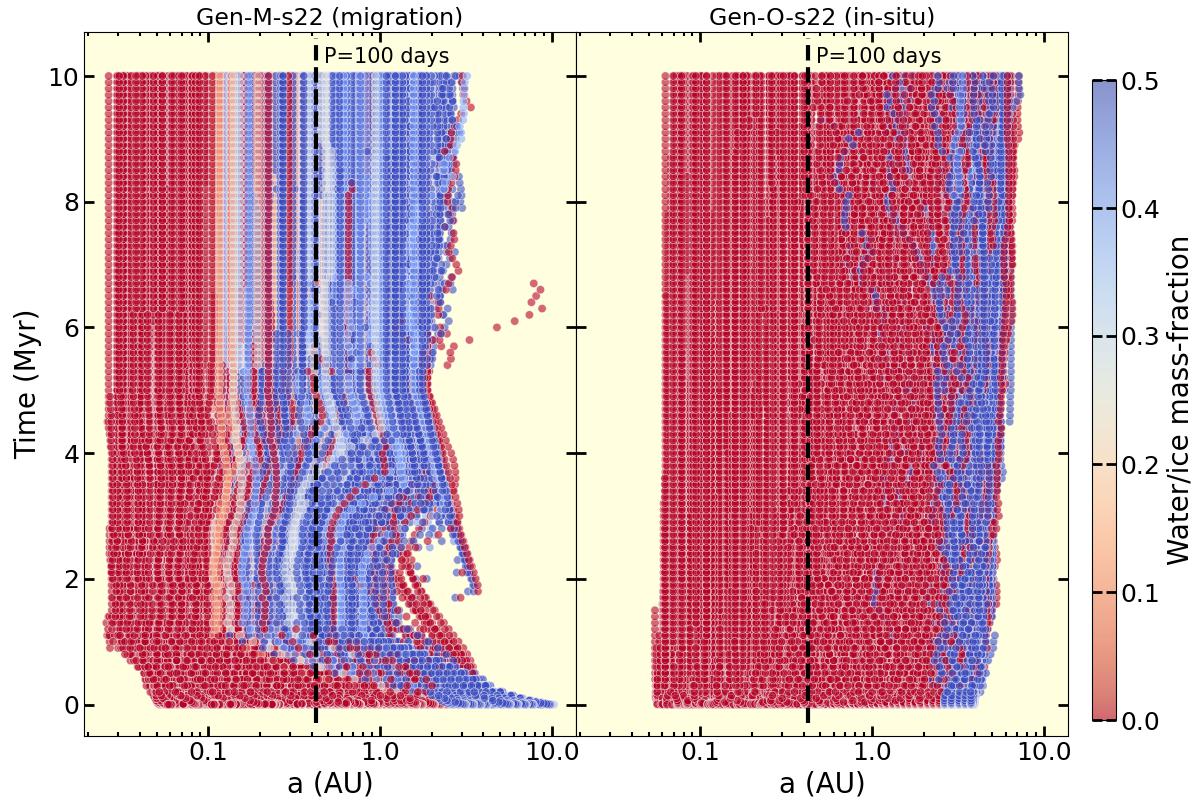}
\caption{Water mass-fractions of the embryos of a migration suite and an in-situ suite of simulations from the Genesis database evolving over the course of simulation, considering a G-type host star. Migration models can explain the transport of water-rich cores to the observable region i.e., within orbital periods of 100 days, while the in-situ models cannot.}
\label{fig:wmf-evol}
\end{figure}

\begin{figure}
\centering
\includegraphics[scale=0.5]{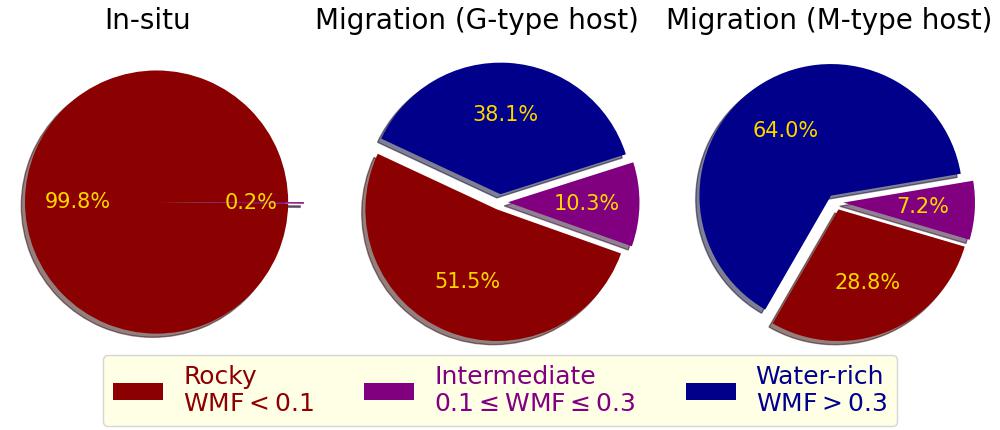}
\caption{Distributions of water mass-fraction (WMF) of the cores produced by all the migration models combined and the in-situ models combined from the Genesis database. Note that, the fraction of cores with intermediate WMF is low; the majority is either predominantly rocky or with high water mass-fraction ($>0.3$).}
\label{fig:wmf-pie}
\end{figure}

\begin{figure}
\centering
\includegraphics[scale=0.4]{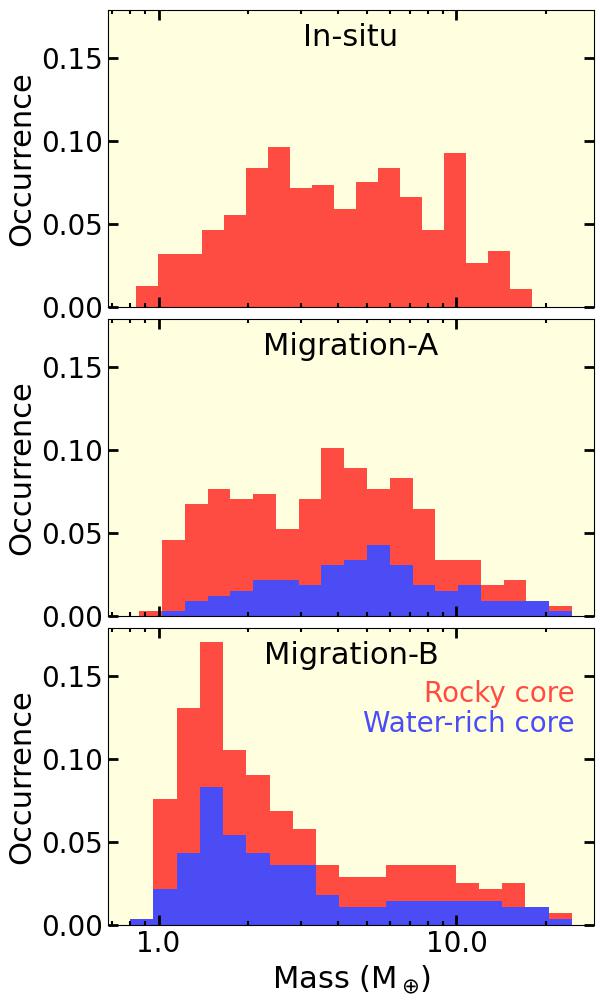}
\caption{Distributions of the core mass of the planets we adopted from the Genesis database segregated as rocky cores and water-ice-rich cores by considering a G-type host star. Note that, in our migration models, there is no clear separation in mass between the rocky cores and the water-rich cores which shows why compositional dichotomy is unlikely to work in our models.}
\label{fig:core-dist}
\end{figure}

\subsection{The Cores and H/He Atmospheres of the Genesis Planets}\label{sec:meth/coreatm}
The final states of simulations of the Genesis planets provide us with the distributions of mass and semi-major axis of the planetary cores. We choose the cores with orbital period ($P$) $< 100$ days and size within 1 \re ~and 4 \re~ as we are interested in the super-Earth and sub-Neptune populations. We consider two kinds of bulk compositions of the cores:
\begin{itemize}
\item Earth-like rocky composition with 32.5\% Fe and 67.5\% MgSiO$_{\rm 3}$ by mass, as the detailed works on mass-radius-relations of the observed terrestrial planets indicate the dominant presence of such cores \citep[][etc.]{zeng16, dressing15, carter12}.
\item Earth-like rocky composition (50-100\% by mass) + a layer of H$_{\rm 2}$O (0-50\% by mass).
\end{itemize}
To calculate the water/ice mass-fraction (WMF) of the cores, we assume that, in the beginning of the simulations, the embryos/planetesimals within the snowlines are fully rocky (100\% Earth-like + 0\% H$_{\rm 2}$O) and beyond the snowlines are water-ice rich (50\% Earth-like + 50\% H$_{\rm 2}$O) \citep{raymond18a,izidoro21}. Over the course of the simulations, the water mass-fractions are evolved in post-process by updating them after each collision according to their mass-ratios and pre-collision water mass-fractions. The evolution of the water mass-fractions for a migration model (Gen-M-s22) and an in-situ model (Gen-O-s22) from the Genesis database \citep{mulders20} is shown in Figure~\ref{fig:wmf-evol}. We identify the planets with water mass-fraction $> 0.3$ as a water-rich planet which is primarily motivated from observations as explained in Section~\ref{sec:ww/def}. The relative abundance of the rocky planets (WMF $<0.1$), water-rich planets (WMF $>0.3$) and planets with intermediate water-rock composition ($0.1<$ WMF $<0.3$) are shown in Figure~\ref{fig:wmf-pie}. To assess the impact of our assumption of WMF being a one-step function of the distance from the host star, we repeated our calculations with a two-step function. We did not find a significant change in the relative abundances of the rocky, water-rich and intermediate planets. Hence, we proceed with the results from our one-step approximation as mentioned above.

Figure~\ref{fig:wmf-pie} shows a significantly low abundance of the planets with intermediate water-rock composition from the migration models which is consistent with our definition. Also, the fraction of water-ice-rich cores is much higher around M-type stars than G-type stars due to the fact that the snowline is much closer in the case of M-type stars (see Table~\ref{tab:host}). Figure~\ref{fig:wmf-evol} and \ref{fig:wmf-pie} re-confirm that only migration models can explain the efficient transport of high mass-fraction of H$_{\rm 2}$O to the planets within orbital periods of 100 days.

We consider that all the planets accrete H/He atmospheres from the disk and retain that after the gas-disk dispersal. We consider a log-uniform distribution for the initial atmospheric mass-fraction ($X_i$) of the planets following \cite{rogers23} as: 
\begin{equation}
\log X_i = \unif(10^{-3}, 0.1)
\end{equation}
The upper limit is consistent with the estimations of the atmospheric mass-fractions ($X$) of the observed sub-Neptunes from their mass and radius measurements \citep[e.g.,][]{lopez14, wolfgang15}. It also covers the range of atmospheric mass-fractions possible after the ``boil-off" stage  \citep{rogers23, misener21}. We consider any planet with $X\lesssim 10^{-4}$ as a stripped super-Earth.

\subsection{Converting mass to radius}\label{sec:meth/m2r}

The size of the planetary cores are calculated by using the model of \cite{zeng19}, as:
\begin{equation} \label{eq:mr}
r_c = f({\rm WMF})~m_c^{1/3.7},
\end{equation}
where $r_c$ is the core radius in \re, $m_c$ is the core mass in \me, and $f$ is the mass-radius constant. $f$ is a function of the water mass-fraction (WMF) of the cores and is calculated by following \cite{zeng19}. In the case of water-ice-rich planets, \cite{zeng19} assume that the bulk of \ho exists in the deep interior in solid phase along the liquid-solid phase boundary. This model also includes a thin isothermal vapor/liquid/super-critical fluid envelope on top of the ice layer. Equation~\ref{eq:mr} provides an approximate expression for the overall structure. The distributions of the core masses of the Genesis models are shown in Figure~\ref{fig:core-dist}.

The existing models for calculating the size of a planet having an atmosphere for a given value of $X$ and equilibrium temperature ($T_{eq}$) are discussed by \cite{rogers23}. For a planet with age $>1$ Gyr, we found that the models by \cite{lopez14} and from the publicly available code \textit{evapmass} by \cite{owen20} are in good agreement and both account for the contraction of the atmospheres of the planets due to cooling over time. As pointed out by \cite{rogers23}, the model by \cite{zeng19} considers a temperature parameter which is the temperature at a pressure of 100 bar and can be interpreted as the equilibrium temperature only when the planets are old enough (age $>1$ Gyr). Thus the model by \cite{zeng19} may be used to calculate only the size of the young planets. As we focus on the exoplanets with age $> 1$ Gyr in all our calculations, we consider the model by \cite{lopez14} to compute the size of atmospheres. Also, the size of an atmosphere does not depend on its metallicity as \cite{lopez14} showed that any difference between the solar-metallicity model and the enhanced-opacity model gets erased after several Gyr. The atmospheric mass-fraction of the planets and hence, their sizes further change as they lose their atmospheres over time through different mechanisms explained in the next subsection. These loss-mechanisms shape the final architectures of the planets which can be compared with the observed planetary statistics.

\subsection{Loss of Atmosphere}\label{sec:meth/loss}
The terrestrial planets can lose their atmospheres due to Parker-type winds. We discuss here two of the leading theories of such mass-loss mechanisms: photo-evaporation loss and impact loss. The effect of core-powered mass loss on the size distribution of small exoplanets across orbital periods is essentially similar to that of photo-evaporation \citep[e.g.,][]{rogers21b} and hence, not included in this work. The photo-evaporation loss mechanism has been widely studied \citep{owen20, owen17} and is found to robustly reproduce the bimodal size distribution of the Kepler planets through simulation. Depending on the density and distance from the stars, the planets could entirely or partially lose their atmospheres, resulting in an overall atmospheric dichotomy. On the other hand, the impact theory suggests that the planets undergoing giant impacts (mass-ratio $\gtrsim$ 0.1) can completely lose their primordial atmospheres regardless of their mass and distance from the host stars \citep{izidoro22, biersteker19}, exposing their compositional diversity. Here, we assess if the Genesis planets undergo adequate number of giant impacts to reflect their compositional dichotomy, i.e. rocky versus water-ice-rich composition, on their size distribution. We subject the 3 different models explained in Section~\ref{sec:meth/gen} to these atmospheric mass-loss processes to assess which combinations of models and mass-loss mechanisms turn out to be consistent with the Kepler size distribution.

\begin{figure}
\centering
\includegraphics[scale=0.49]{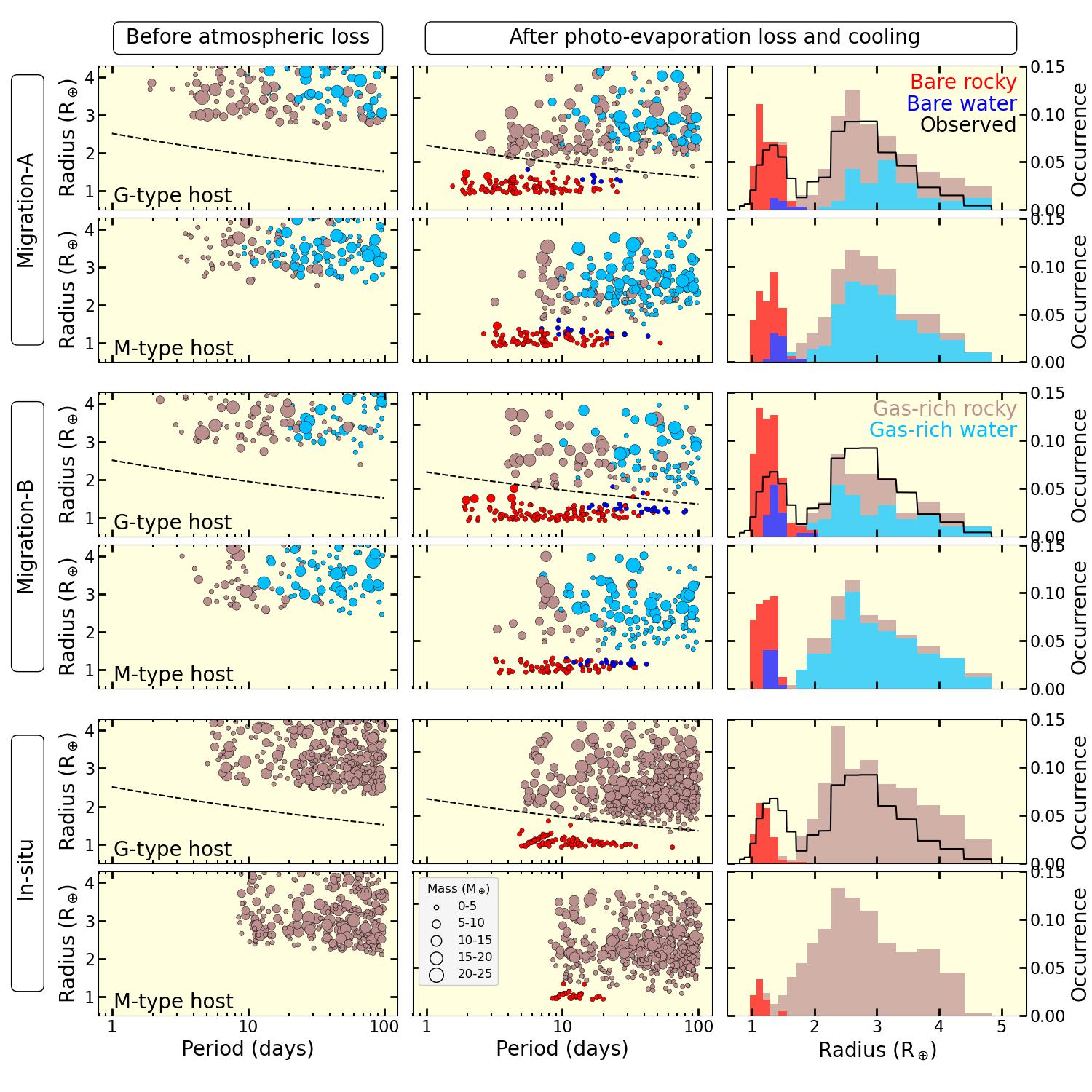}
\caption{The radius of the Genesis planets versus their orbital period without any atmospheric loss (left) and with photo-evaporation loss (middle) and the radius distributions of the planets of all periods after photo-evaporation loss (right). The black dashed lines in the first two columns show the
observed slope of radius valley over orbital period \citep{izidoro22, vaneylen19, gupta19}. The observed radius distribution for the Sun-like (G-type) stars is taken from \cite{fulton18}.}
\label{fig:photev-rp}
\end{figure}

\subsubsection{Photo-evaporation Loss}\label{sec:meth/loss/ph}
We calculate the photo-evaporation mass-loss rate and timescale by following the formalism of \cite{owen17}. The mass-loss rate depends on the high energy luminosity ($L_{HE}$) of the host stars and an efficiency parameter $\eta$ denoting the efficiency of these high energy photons for mass-removal \citep{owen17}. We calculate $L_{HE}$ as a function of mass and age of the host stars by following \cite{owen17} who assume a linear dependence of $L_{HE}$ on the stellar mass. Additionally, we adopt the same saturation timescale of 100 Myr for both types of stars, as proposed by \cite{owen17}. They argue that the time-integrated XUV flux is predominantly influenced by the initial 100 Myr, making it unlikely that a more extended saturation timescale for M-dwarfs would significantly alter the outcomes. Any potential differences in this regard are expected to be accommodated by the range of values assumed for $\eta$. We numerically solve the mass-loss equations and take the size of the planets after an age of 5 Gyr as the final sizes of the planets since they do not change significantly afterwards. 
Following \cite{rogers21, owen17}; etc., we calculate the photo-evaporation efficiency, considering the energy-limited scenario, as a function of the escape velocity of the planets ($v_{esc}$) as:
\begin{equation} \label{eq:eta}
\eta = \eta_0 \left(\frac{v_{esc}}{25~km/s}\right)^\alpha
\end{equation}
$\eta_0$ and $\alpha$ are free parameters of our models and hence, we run our simulations over a grid of values for $\eta_0$ and $\alpha$. This approach helps us address the dependence of $\eta$ on the planet mass. H/He takes longer time to escape from massive planets, allowing them to lose more energy through radiative cooling and slowing down the mass-loss process \citep{owen12}. While energy-limited escape applies to the weakly irradiated planets, some of the planets in our models could lose their atmospheres through recombination-limited escape \citep{owen16, murray-clay09} or X-ray evaporation \citep{owen12}, which would further complicate the situation. However, \cite{owen16} showed that most of the small planets with both mass and radius measurements fall into the energy-limited regime. Moreover, the energy-limited approach adopted in this paper (i.e., Equation~\ref{eq:eta}) has been shown to well reproduce the Kepler size bimodality \citep[e.g.,][]{rogers21, owen17}. For all the model-suites with G-type host stars, we find that the size distribution of the simulated planets, especially the location of the radius valley, turns out to be consistent with that from Kepler when we adopt $\eta_0 < 0.06$ and $0 < \alpha < 0.5$. Since, we do not have an observed size distribution for the planets around late M-dwarfs to date to benchmark with, we use the same values of $\eta_0$ and $\alpha$ for the case of M-type host stars and still find a bimodal size distribution of the simulated planets as speculated \citep{cloutier20}.

Figure~\ref{fig:photev-rp} shows that the migration-A model is found to provide the best match with the Kepler size distribution. However, photo-evaporation can produce the ``evaporation valley" with all the three suites of models and for both the types of host stars. It's the fractional occurrence of water-ice-rich planets that distinguishes the three models.

\begin{figure}
\centering
\includegraphics[scale=0.49]{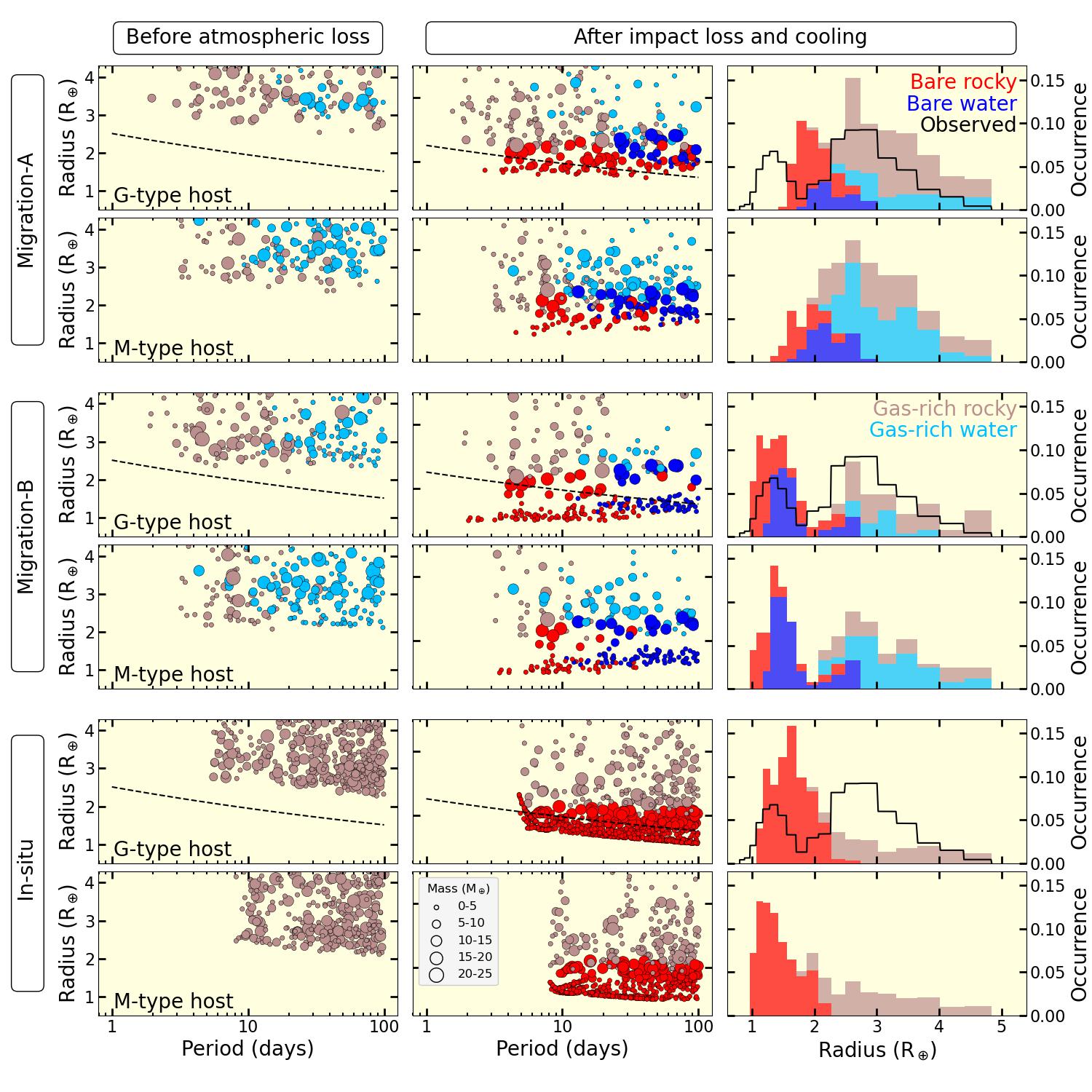}
\caption{Same as Figure~\ref{fig:photev-rp} but considering only impact loss. Evidently, impact loss of H/He atmospheres is less efficient than photo-evaporation loss in creating the radius valley in our models.}
\label{fig:impact-rp}
\end{figure}

\subsubsection{Impact Loss}\label{sec:meth/loss/im}
Giant impacts during planetary accretion can deliver significant energy to the planet, thereby heating the core and the envelope, which can cause hydrodynamic escape of H/He envelope leading to a rapid mass loss. The effect of the shockwave generated by a giant impact can also eject a fraction of the envelope, but that effect is less significant compared to thermal effect \cite{biersteker19} and hence, we only consider the latter in this work. Following \cite{izidoro22, biersteker19}, we assume that, after the gas disk dispersal, thermal effect from a giant impact on a planet (where mass of the impacting body $\geq 0.1$ times the planet mass) can completely strip the primordial atmosphere. In the absence of the gas disk, this loss cannot be replenished. The in-situ simulations are performed without any gas disk throughout whereas in the case of migration models, the gas disk is dispersed after 5 Myr from the start of the simulations. We assume that an impact by a less massive planetesimal doesn't change the atmospheric mass-fraction of the planets. 

Figure~\ref{fig:impact-rp} shows that the impact stripping process can somewhat reproduce the size bimodality with migration-B model whereas it absolutely fails to produce the radius valley with migration-A and in-situ models. Moreover, in the case of migration-B, it is not the dichotomy in the bulk composition but rather the dichotomy of atmospheres (presence and absence of atmospheres) that causes the bimodality of the size distribution in our model, similar to the case of photo-evaporation loss, as evident from Figure~\ref{fig:impact-rp}.

\subsection{Discussions on Our Modeling Steps}\label{sec:meth/discuss}

Our study shows that the current suites of migration models of the Genesis database perform significantly better than the in-situ models in reproducing the Kepler size-bimodality. While photo-evaporation is found to robustly explain the size bimodality with different distributions of core mass and location, impact stripping process appears to do so only with certain distributions.

As we aim to combine multiple complex processes with the N-body simulations of the Genesis database we adopt simplified models with certain assumptions at various steps. A few of these simplifications and assumptions are described here which can be seen as a guide for future improvements. First, our models simulate zero to a few ultra-close-in super-Earths with orbital period $\lesssim4$ days, failing to explain the Neptune desert. We plan to address this in our future studies with new N-body simulations with refined initial parameters.  Second, we employ a simple model for the impact stripping mechanism by following \cite{izidoro22} where we assume that lower-mass impactors below the cutoff limit do not cause any erosion of atmosphere. While it is true for a single impact as the plume-driven local mass loss due to an impacting planetesimal may not be significant but the cumulative effect of multiple small impacts can cause significant erosion of atmosphere from the embryos \citep{chen22, schlichting15}. However, this process is unlikely to significantly impact the planetary statistics in our in-situ suite of models. Giant impacts are already observed to completely erode the primordial atmospheres of the majority of planets in this suite, resulting in an over-abundance of bare super-Earths that is inconsistent with the Kepler distribution (refer to Figure~\ref{fig:impact-rp}). Conversely, while giant impacts are found to be effective in explaining the radius valley for the migration-B model, they appear to be inefficient in eroding the atmospheres of planets in the Migration-A model. Therefore, introducing planetesimal-driven escape into the Migration-A model could offer an explanation for the radius valley. However, migrating planets could capture most of the planetesimals without losing their atmospheres before the dispersal of the gas disk and by the time the gas disk disperses, the number of planetesimals may not be large enough to cause significant atmospheric erosion. Nonetheless, we plan to include the effects of planetesimals in the case of migration and a detailed model of plume-driven loss due to impacts by planetesimals in our follow-up studies. Finally, for the water worlds, we adopt the models of \cite{zeng19} who assume isothermal vapor/liquid/super-critical fluid envelopes. Using updated models of water worlds, for example, the adiabatic envelope models of \cite{aguichine21}, can significantly alter the population statistics. Nonetheless, this initial approach to developing population synthesis models based on the Genesis formation models helps us gain essential insights. 

The primary aim of studying the impact stripping process was to test the hypothesis of compositional dichotomy through impact stripping of migrating planets, as proposed by \citep{izidoro22}. However, as we have a different model setup, we find that, impact stripping of the planets from our migration models does not result in compositional dichotomy, but rather accounts for the size bimodality with atmospheric dichotomy, similar to the case of photo-evaporation. This is primarily because the masses of the rocky and water-rich cores from the migration models do not show any bimodality (see Figure \ref{fig:core-dist}) unlike the planets of \cite{izidoro22}, thereby lacking any inherent bimodality essential for compositional dichotomy to work. Additionally, we find that our migration models produce a significant number of bare water-ice-rich planets adding to the super-Earth population in contrast to \cite{izidoro22}. This happens because no small water-ice-rich planet goes through any giant impact in the models of \cite{izidoro22}, whereas such planets are found to lose their atmospheres by both giant impacts and photo-evaporation in our migration models. We use the occurrence of these bare water worlds as a diagnostic to our analysis as explained in the following section. Since the in-situ models do not favor the water-world hypothesis, we assess the possibility of the water worlds in the following sections with the help of our migration models.

\section{The Water Worlds: From Models and Observations}\label{sec:ww}
As migration models suggest the possibility of water-ice-rich super-Earths and sub-Neptunes, we look for similar patterns from observations. We consider the planets with both mass and radius measurements for this. Such planets, albeit currently small in sample size, allows us to look into the mass-radius-orbital period distributions of the planets that can be compared with model outcomes.

\subsection{The Sample List of Luque and Palle 2022 and the TEPCat Catalogue}\label{sec:ww/lptep}
\begin{figure}
\centering
\includegraphics[scale=0.5]{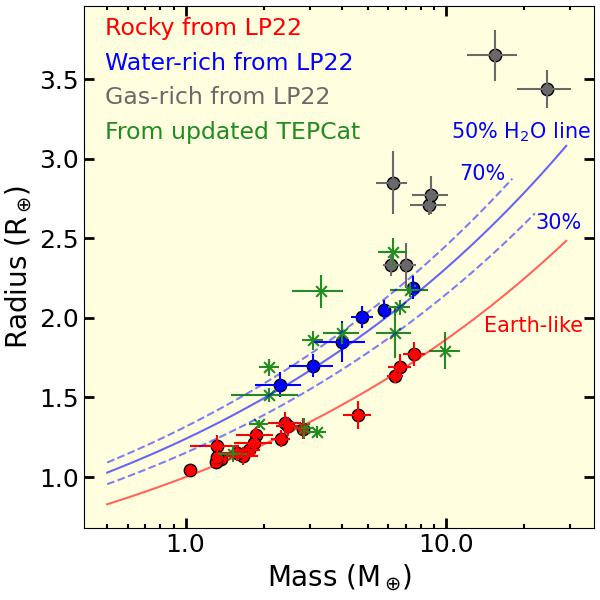}
\caption{The mass-radius relations of the planets from LP22 and the latest updated TEPCat \citep{southworth11}.}
\label{fig:lp-tep}
\end{figure}

\cite{luque22} (hereafter, LP22) suggested three populations of planets: ``rocky", ``water-rich", and ``gas-rich" from  the mass-radius relations and the density distribution of the planets around M-dwarfs that they studied. Although the density of a ``water-rich" planet can also be explained with a rocky planet having a thin layer of atmosphere, it is the distinct clustering of the ``water-rich" planets in the mass-radius-density space that interests us as we find similar planets from our migration models. Even after updating the sample list by including planets from the latest updated Transiting Extrasolar Planets catalogue\footnote{\url{https://www.astro.keele.ac.uk/jkt/tepcat/}} (TEPCat, \cite{southworth11}) around M-dwarfs, we find a similar pattern. Although a few planets have now appeared with intermediate densities (see the green points on Figure~\ref{fig:lp-tep}), the fraction of such ``water-rich" planets is found to be almost unchanged which motivates us to use this sample list to define the water worlds.

\subsection{Revised Definition of Water Planets}\label{sec:ww/def}
We assume that the ``water-rich" planets identified by LP22 are truly water planets and verify if their fractional occurrence is consistent with our model predictions. This would imply that these planets have significantly lost their primordial H/He atmospheres and hence, we call them bare water planets (BWP). On the other hand, our migration models suggest that a fraction of the gas-rich planets could also be made up of water-ice-rich cores which we call the gas-rich water planets (GWP). Following LP22, we define the BWPs as the planets strictly falling on or around the 50\% \ho line in the mass-radius diagram. Keeping in mind the uncertainty in the models of the mass-radius relations, we allow a small range of water mass-fractions around 0.5 for the definition of BWPs. To accommodate the ``water-rich" planets of LP22 into our definition, we set this range between the 40\% \ho line and the 70\% \ho line in the mass-radius diagram (see Figure~\ref{fig:lp-tep}). To ensure this, we introduce a new parameter $\f$ which is somewhat equivalent to the mass-radius constant ($f$) in Equation~\ref{eq:mr} as:
\begin{equation} \label{eq:fp}
\f = \frac{r_p}{m_p^{1/3.7}},
\end{equation}

where $m_p$ and $r_p$ are the observed mass and radius of the planets in Earth units. For a bare planet, $\f$ becomes exactly equal to $f$. Thus for a bare Earth-like rocky planet, $\f$=$f(0)$=1, and according to our definition, BWPs are the planets with $\f$ between $f(0.4)$=1.1976 and $f(0.7)$=1.3164. Following this, we identify the planets with $F \geq f(0.7)$ as gas-rich planets. Accordingly, we also identify the planets from our models as water planets (bare or gas-rich) if their water fractions are $\geq 0.4$. This constraint is consistent with the models as the water fractions of most of the planets from our models are either 0-0.1 or 0.3-0.5 (see Figure~\ref{fig:wmf-pie}). This is also justified from the observational perspective as the water planets with much lower water fractions would be difficult to detect in the initial attempts. The upper limit of 0.7 on water fraction is also consistent with our models as none of the planets from our models have water fraction $>0.5$ and most of the gas-rich planets are found with $\f > f(0.7)$. 

We use this definition to identify the bare rocky, bare water, and gas-rich planets from observations and from our models. For observed planets around M-dwarfs, we use our updated list mentioned in Section~\ref{sec:ww/lptep} and in the case of G-dwarfs, we use the planets from the latest version of TEPCat. We apply the same constraint on the precision of mass and radius of the planets as LP22 in all cases, i.e, 8\% on radius and 25\% on mass, with an exception in the case of Kepler-138 d. We include Kepler-138 d in our sample list of planets around M-dwarfs, albeit having a mass uncertainty of $\sim$33\%, as it is a strong water-world candidate suggested by previous studies \citep{piaulet23}.  We then compare the fractional occurrence of the bare water planets predicted by our models with that from observations.

\begin{figure}
\centering
\includegraphics[scale=0.5]{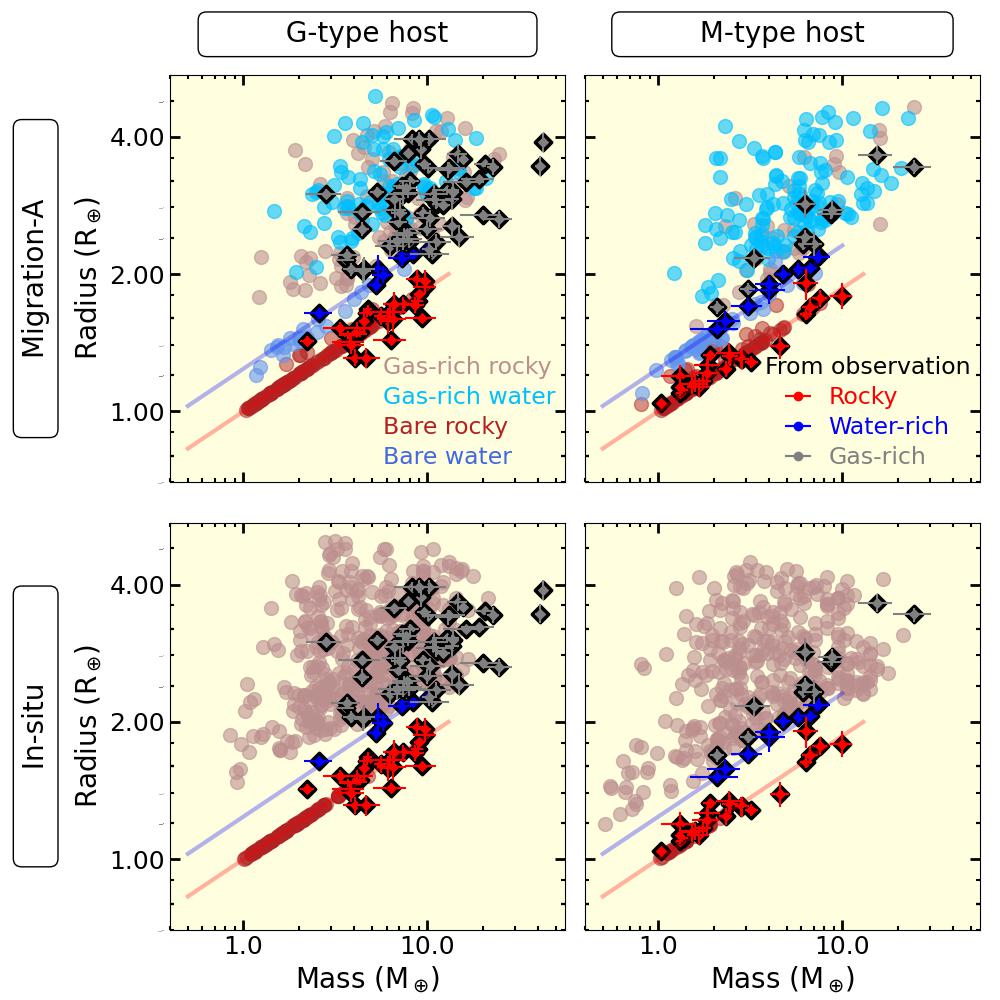}
\caption{The mass-radius relations of the simulated planets from our migration and in-situ models both of which are found to explain the mass-radius relations of the observed planets. However, they predict completely different bulk compositions which can be reflected in the atmospheric composition.}
\label{fig:mr}
\end{figure}

\begin{figure}[!ht]
\centering
\includegraphics[scale=0.45]{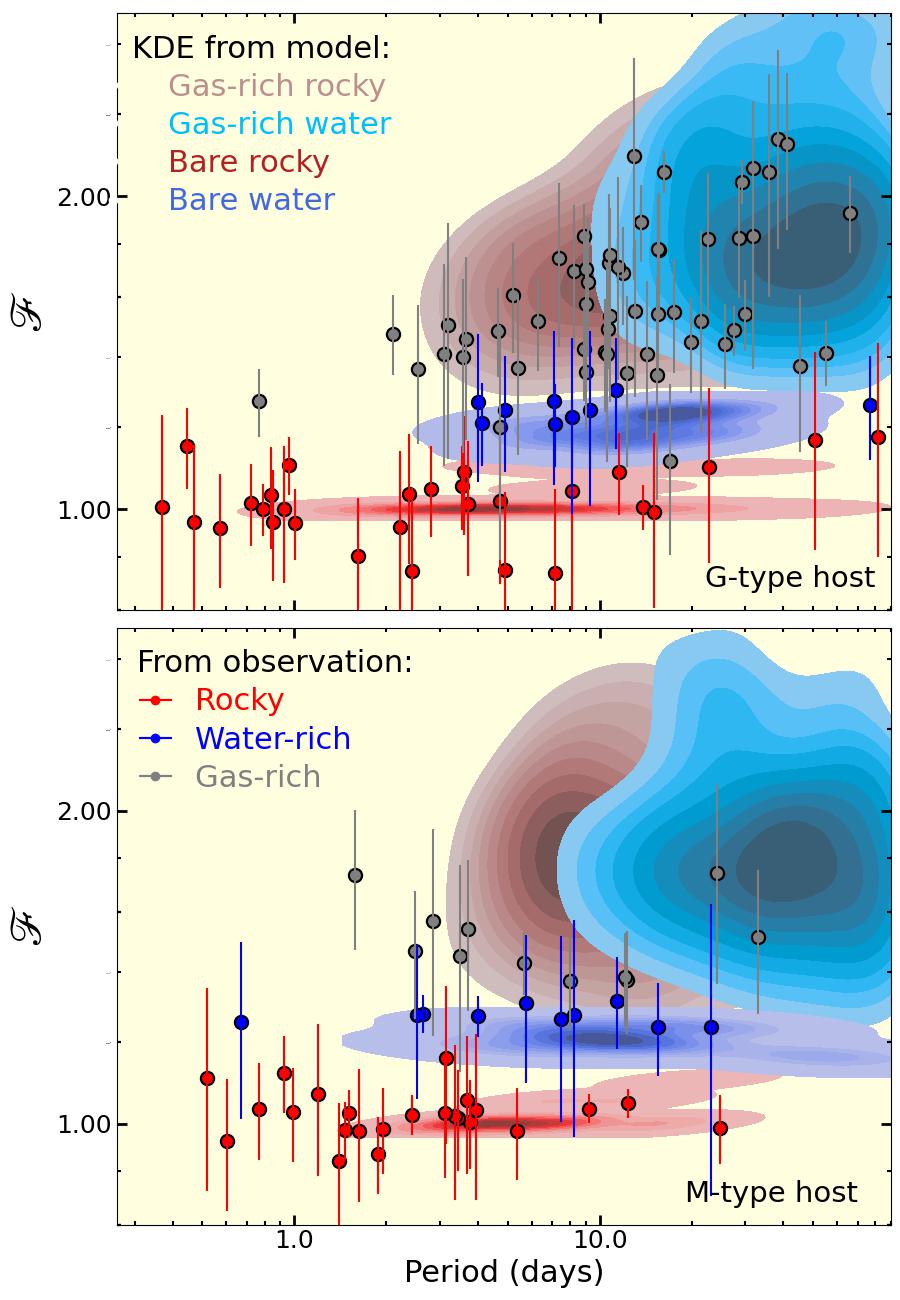}
\caption{The kernel density estimation (KDE) of the occurrences of the simulated planets in $\f$-period space calculated separately for the bare rocky, bare water, gas-rich rocky and gas-rich water planets. The KDEs can be used to calculate the probability density functions over the entire $\f$-period space. Likely natures of the observed planets identified only from their $\f$ values are found to be consistent with the background KDEs.}
\label{fig:kde}
\end{figure}

\begin{figure}[!ht]
\centering
\includegraphics[scale=0.45]{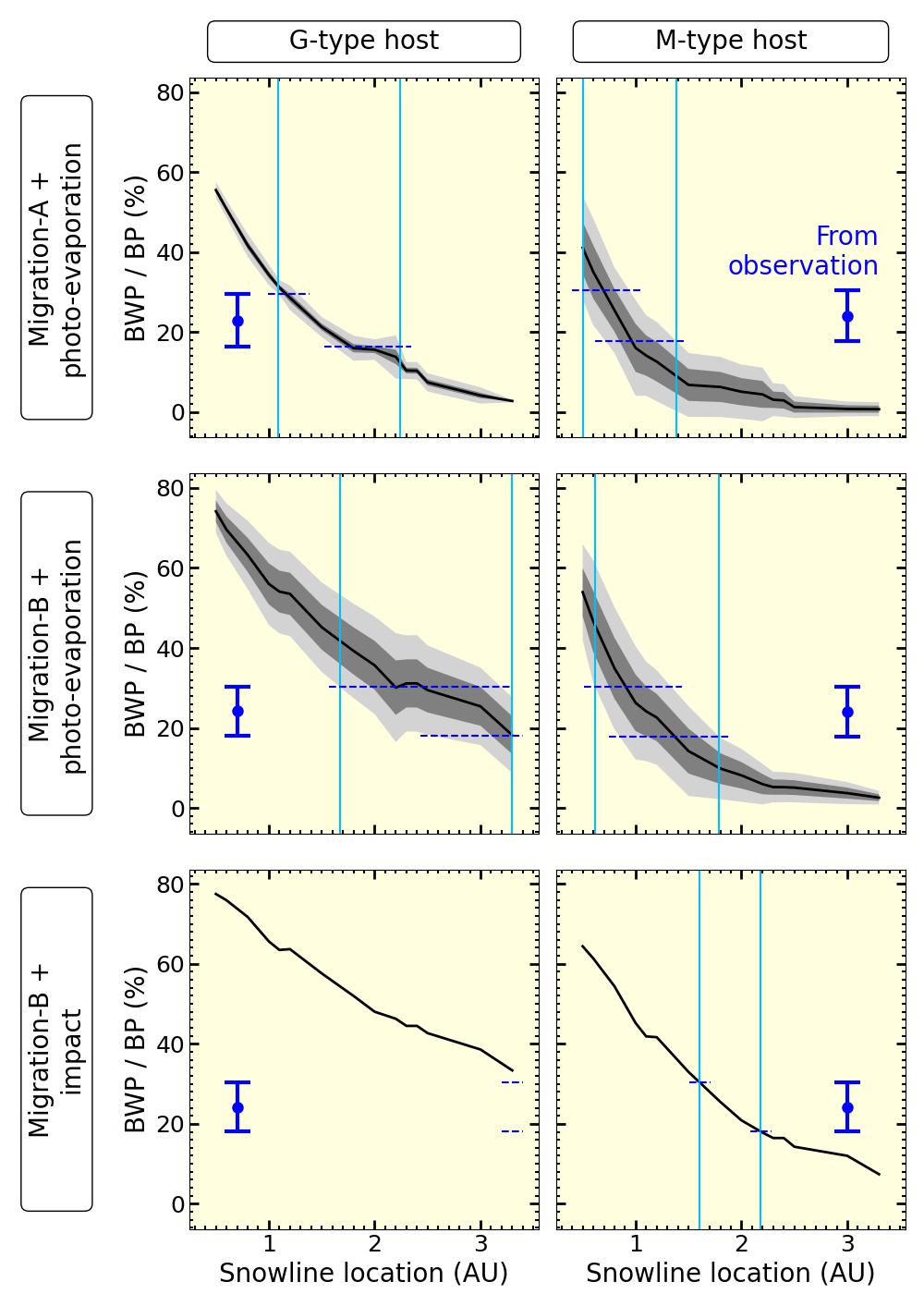}
\caption{Variation of the percentage occurrence of the bare water planets among the bare planets (BWP/BP) calculated from our models by using their period-$\f$ KDEs with the location of snowline of the disk. The dark- and light-shaded regions denote the 1$\sigma$ and 3$\sigma$ uncertainties respectively. The blue errorbar denotes the 1$\sigma$ uncertainty range of the BWP/BP occurrence computed from observations. The cyan vertical lines denote the possible range of snowline locations that can explain the upper limit of the occurrence of the bare water planets derived from observations. Impact loss is very efficient at stripping H/He atmospheres, thereby creating a lot of bare water planets. As observations indicate much lower occurrence (upper limit) of bare water worlds (see Table~\ref{tab:bwpfrac}), this alternatively requires the snowline to be further away.}
\label{fig:sl}
\end{figure}

\begin{deluxetable}{cCCCCc}[!h]
\tablecaption{The maximum possible occurrence of bare water planets as a fraction of bare planets (BWP/BP) from observation and model.\label{tab:bwpfrac}}
\tablewidth{0pt}
\tablehead{
\colhead{Host} & \colhead{From} & \colhead{Migration-A} & \colhead{Migration-B} & \colhead{Migration-B} & \colhead{In-situ} \\
 \colhead{star} & \colhead{observation}  &  \colhead{+ photo-evap} &  \colhead{+ photo-evap} & \colhead{+ impact}      & \colhead{}
}
\startdata
G         & <24.2\pm6.1\%   & 14.8\pm1.8\%    & 30.1\pm6.8  & 46.3\pm0\%  & 0\%  \\
M         & <23.3\pm5.7\%   & 25.5\pm5.3\%   &  34.9\pm7.6  & 54.4\pm0\%  & 0\%  \\ 
\enddata
\tablecomments{The uncertainties in the upper limits obtained from observation appear from the uncertainties in the mass and radius of the planets. The uncertainties in the model values appear from the random distribution of $X_i$ and also from the range of values adopted for $\eta$ in the case of photo-evaporation. The snowline locations are chosen at 2.2 AU and 0.8 AU for the G-type and M-type stars respectively.}
\end{deluxetable}

\subsection{Mass-Radius-Period Distributions and Occurrence of Water Planets}\label{sec:ww/occ}
Both migration and in-situ models are found to explain the observed mass-radius distributions for both the types of host stars, as evident from Figure~\ref{fig:mr}. This is also on par with the arguments of \cite{rogers23}. Therefore, although current observations cannot be used to draw any inference about the water worlds, we leverage our migration models to prescribe a systematic set of diagnostics that can motivate future observations.

We calculate the maximum possible occurrence of the bare water planets as a fraction of the bare planets from current observations by using the definition described in Section~\ref{sec:ww/def}, considering the error-bars of the masses and radii of those planets. Note that, these values only represent the upper limits of the occurrence of bare water planets as the possibility of them being rocky planets with thin atmospheres cannot be ruled out. Table~\ref{tab:bwpfrac} shows the percentage occurrence of the BWPs among bare planets calculated from observations (upper limits) and from our migration models for both G- and M-type host stars. We choose 3 of the combinations of migration models and mass-loss mechanisms mentioned in Section~\ref{sec:meth/loss}: migration-A + photo-evaporation, migration-B + photo-evaporation, and migration-B + impact stripping. To predict the occurrence of the water planets from our models, we utilize the $\f$-period distributions of the simulated planets which is described as follows. We first compute the kernel density estimations (KDE) from the $\f$-period distributions of the simulated planets separately for the four types of compositions, viz., bare rocky, bare water, gas-rich rocky, and gas-rich water, as shown in Figure~\ref{fig:kde}. The KDEs essentially represent the conditional probability density functions (PDF) of finding a planet of these given compositions at a particular point on the $\f$-period plane. Subsequently, we integrate the PDFs across the range of $\f$ values and orbital periods covered by the samples of observed planets and calculate the probability that a randomly selected bare planet from that plane is a water world. Evidently, KDEs calculated from our models are found to be consistent with our compositional classification of the observed planets. The occurrences shown in Table~\ref{tab:bwpfrac} correspond to our fiducial values of the locations of snowline of the disks (see Table~\ref{tab:host}).

We also calculate the occurrences for other values of the locations of the snowline. Figure~\ref{fig:sl} shows which range of values of the locations of snowline is consistent with observations. Clearly, the fraction of bare water planets from the photo-evaporation models are consistent with the snowline locations suggested by the disk models \citep{raymond07, mulders15}. Conversely, the same produced by the impact models are found to be inconsistent with the upper limits obtained from observations, which otherwise suggests the snowline locations to be unusually away from the host stars ($>3$ AU for G-type host and $\gtrsim1.6$ AU for M-type host).

\subsection{Discussions on the Possibility of Water Worlds}\label{sec:ww/discuss}
Our study shows that the migration + photo-evaporation models strongly favor the water world hypothesis. The impact stripping mechanism is found to produce more number of water planets compared to photo-evaporation when subjected to the same distribution of the planets. Overall, both the migration models can well explain the upper limits set on the occurrence of the bare water planets derived from observations with some adjustments with the models and the locations of snowline. Moreover, Figure~\ref{fig:kde} shows that the KDEs can be used to estimate the likelihood of a gas-rich planet having a water-rich core which otherwise can not be inferred directly from observations.

Table~\ref{tab:bwpfrac} shows that our migration models tend to produce more fraction of bare water planets around M-dwarfs than around G-dwarfs. In contrary, observations suggest more of a similar BWP occurrence for both the types of host stars. This questions the validity of the water world hypothesis. Alternatively, this discrepancy might arise because previous spectroscopic missions did not look where migration models suggest the highest concentration of water worlds.

\begin{figure}
\centering
\includegraphics[scale=0.38]{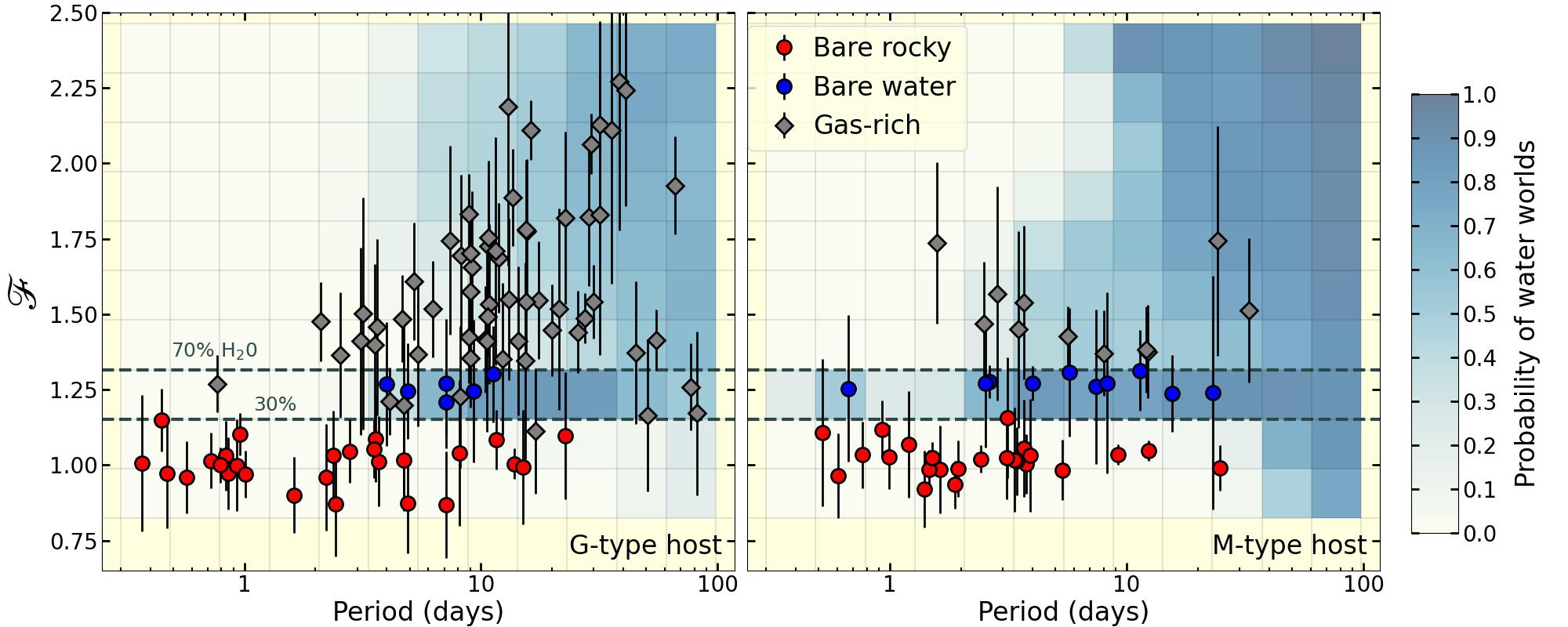}
\caption{Color-map of probability of a planet with a given orbital period and $\f$ value being a  water world as predicted by our migration+photo-evaporation models. The overplotted error-bars denote the observed planets from our sample lists and their colors denote the likely nature based on their $\f$ values only. Evidently, the planets around M-dwarfs are more likely to be water planets than those around G-dwarfs, especially at longer orbital periods.}
\label{fig:wwprob2d}
\end{figure}

\section{Guidance to Future Search for Water Worlds} \label{sec:guide}
\begin{figure}[]
\centering
\includegraphics[scale=0.4]{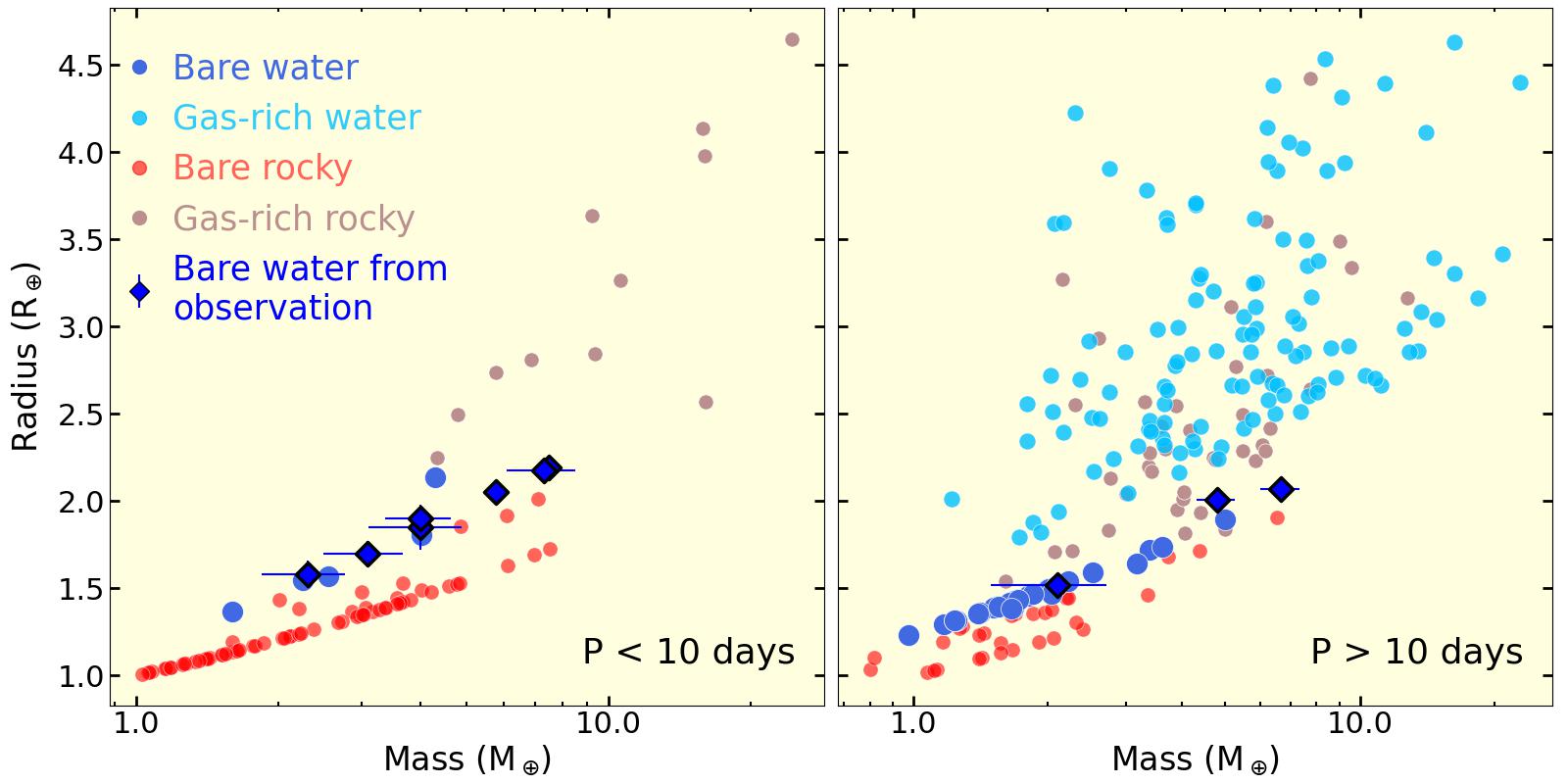}
\caption{Mass-radius distributions of the simulated planets from our migration + photo-evaporation model around G-type host stars showcasing that the water-rich planets are abundant beyond orbital period of 10 days. The water-rich planets are enlarged for clarity.}
\label{fig:wwmr}
\end{figure}

While the predicted fraction of bare water planets is consistent with current observed exoplanets, these water worlds are not equally distributed across the parameter space. We calculate the likelihood (2D PDF) of detecting a water-ice-rich planet over a rectangular grid of orbital periods and $\f$ values from our migration models with the help of the same KDEs mentioned in Section~\ref{sec:ww/occ}. Figure~\ref{fig:wwprob2d} shows the corresponding map of the likelihood of occurrence of a water-rich planet (either bare or gas-rich) over the $\f$-period plane. Observed planets are overplotted in the figure to show the probabilities of the bare and gas-rich planets identified from observations being water worlds. Our study shows a strong dependence of the likelihood of water planets on the orbital periods. We show the mass-radius distributions of the simulated planets from the migration-A + photo-evaporation model around M-type host stars in Figure~\ref{fig:wwmr} for two different ranges of orbital periods: $P < 10$ days and $P > 10$ days. Clearly, the water worlds are more abundant beyond orbital periods of 10 days, implying that the snowline has effectively moved from $\sim$0.8 AU to $\sim$0.06 AU (10 days). This is slightly longer than the range over which most planets have been followed up by spectroscopic observations to date. We find a similar pattern around the G-type hosts stars. 

\begin{figure}[]
\centering
\includegraphics[scale=0.5]{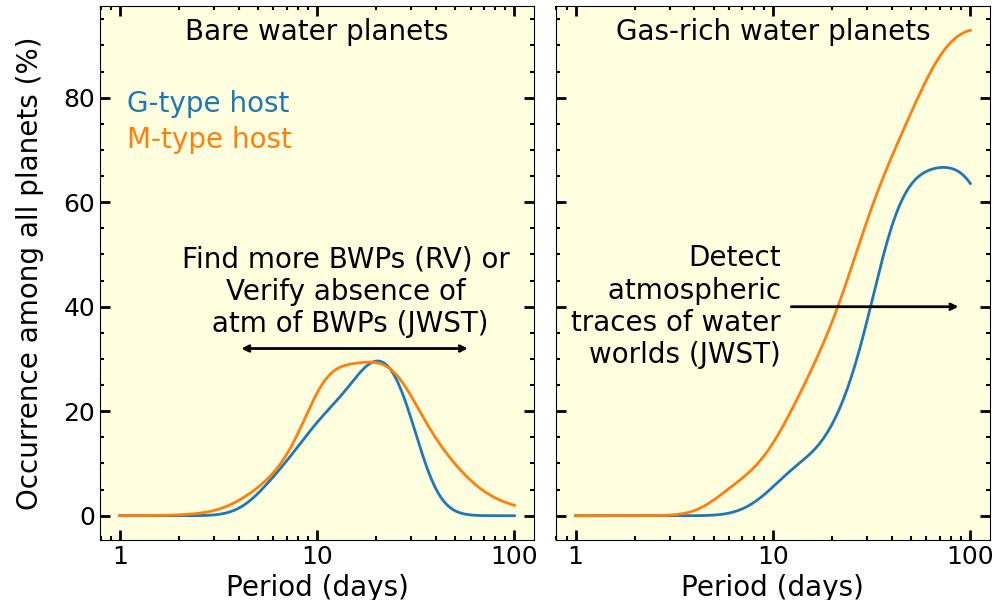}
\caption{Fractional occurrence of the water worlds as functions of the orbital period predicted by our migration + photo-evaporation model for the G-type host star. While the bare and gas-rich water planets are likely to be concentrated beyond the orbital periods of 10 days, it is more likely to find a water planet around an M-dwarf than around a G-dwarf.}
\label{fig:wwprob1d}
\end{figure}

The period-dependence of the occurrence of water worlds is calculated by computing the 1D PDF of the occurrence of the bare and gas-rich water planets as a function of orbital period which is shown in Figure~\ref{fig:wwprob1d}. As evident from the figure, the PDFs for both the bare and gas-rich water planets peak beyond orbital periods of 10 days. The total likelihood (area under the curve) of finding a bare or gas-rich water planet is also high beyond orbital periods of 10 days. The occurrence of bare water planets drops at $\gtrsim 50$ days due to the over-abundance of gas-rich planets. Hence, future radial velocity (RV) follow-up efforts could prioritize longer orbital periods (10-50 days) to increase their odds of finding bare water planets. Also, since the absence of an atmosphere on such a planet could strongly indicate a water-rich core, this is the range of orbital periods where spectroscopic or phase-curve studies \citep{kempton23} with JWST and future space-bound and ground-based missions would have a higher probability of identifying water worlds.

On the other hand, we find from our models that the likelihood of finding a gas-rich water planet increases as we go further away from the central star. Thus JWST and future atmospheric survey missions should look beyond the orbital periods of 10 days to verify the existence of the water-rich sub-Neptunes. Atmospheric features that can be used as traits for their water-ice-rich bulk composition are high mean molecular weight or high metallicity \citep{kempton23} of the atmospheres, and high abundance of water vapor in the atmospheres, among others. Table~\ref{tab:jwst} shows a list of shortlisted planets with known mass and radius for which the calculated probability of being water worlds from our migration models is $\gtrsim$50\%. The table also shows the values of their transmission spectroscopic metric (TSM) for observation using the JWST/NIRISS instrument \citep{kempton18}. The uncertainties in the probabilities include the effect of the uncertainties in the model parameters (e.g., photo-evaporation efficiency) but are dominated by the uncertainties in the measured mass and radius of the planets, especially for the planets with intermediate $\f$ values (likely bare water planets). While it is currently difficult to improve the precision of the planetary radii owing to the limits posed by the precision in the stellar radii, future RV studies could focus on improving the precision in their mass measurements.

\subsection{Discussions on the Water-World Candidates}\label{sec:guide/discuss}
Table~\ref{tab:jwst} shows the list of potential water-world candidates which could be followed up by JWST to find atmospheric tracers of their volatile contents. This list contains some of the targets that are already suspected to be water worlds on the basis of observations with JWST or HST. For example, K2-18 b, which is suggested to be a gas-rich water world by our models, is suspected to be a hycean (hydrogen + ocean) planet from recent observation with JWST \citep{madhusudhan23}. The observed spectra suggest the presence of a shallow H$_{\rm 2}$ atmosphere which then requires a \ho layer (most likely, in an ocean form) that can explain the bulk density \citep{madhusudhan23}. Again, \cite{piaulet23} claim Kepler-138 d to be a bare water world ($\sim$51\% water by volume) from their interior modeling of the planet which is in strong agreement with our model prediction. Their calculations were further supported by the flat optical/IR transmission spectrum that they obtained from HST, making it a prime target for JWST. 

Conversely, our models suggest that TOI 1695 b, once thought to be a water world based on its mass and radius measurements \citep{cherubim23}, is less likely to be a water world due to its relatively high density and short orbital period. However, that does not rule out the possibility of a low bulk content of \ho as we have restricted our definition of water worlds to a lower limit of 30\% of water mass-fraction. Again, TESS and HST observations of TOI-270 d \citep{mikalevans23} tend to support a  H$_{\rm 2}$-rich atmosphere that shows strong absorption features of \ho. Although this is in disagreement with our models suggesting it to be a bare water planet, it could still be a gas-rich water planet, which requires further follow-up with JWST for confirmation.

\begin{deluxetable}{ccccccc}[!h]
\tablecaption{List of planets with high mean probability (ww prob) of being a water world ($>40$\% for bare and $>60$\% for gas-rich)}
\label{tab:jwst}
\tablewidth{0pt}
\tablehead{
\colhead{Planet} & \colhead{Host type} & \colhead{Nature} & \colhead{Period (days)} & \colhead{ww prob (\%)\tablenotemark{$\dagger$}} & \colhead{TSM} & \colhead{K (m/s)}
}
\startdata
Kepler-1705 c & G & bare & $11.28 \pm 0.0010$ & $46.0 \pm 29.2$ & 2.8 & -- \\
Kepler-138 d & M & bare & $23.09 \pm 0.0006$ & $60.1 \pm 28.0$ & 22.0 & $0.395 \pm 0.09$ \\
TOI-270 d & M & bare & $11.38 \pm 4.61e-05$ & $49.5 \pm 23.0$ & 87.1 & $2.56 \pm 0.23$ \\
TOI-1468 c & M & bare & $15.53 \pm 3.4e-05$ & $54.7 \pm 29.8$ & 64.6 & $3.48 \pm 0.35$ \\
K2-314 d & G & gas-rich & $35.75 \pm 0.0050$ & $61.8 \pm 9.8$ & 18.1 & $1.97_{-0.47}^{+0.54}$ \\
Kepler-33 e & G & gas-rich & $31.79 \pm 0.0002$ & $61.3 \pm 8.8$ & 9.5 & -- \\
Kepler-33 f & G & gas-rich & $41.03 \pm 0.0002$ & $64.0 \pm 8.4$ & 9.8 & -- \\
Kepler-289 c & G & gas-rich & $66.03 \pm 0.0008$ & $66.3 \pm 6.0$ & 12.7 & -- \\
K2-18 b & M & gas-rich & $32.94 \pm 0.0011$ & $79.2 \pm 7.9$ & 41.8 & $3.36 \pm 0.64$
\enddata
\tablenotetext{\dagger}{The uncertainties in ww prob predominantly source from the uncertainties in the radius and mass of the planets.}
\tablecomments{Probabilities of being water worlds (ww prob) are calculated with the help of our migration-A + photo-evapoartion models. The transmission spectroscopic metric (TSM) values are calculated by following \cite{kempton18}.\\
(This table is available in its entirety in machine-readable form.)}
\end{deluxetable}

\clearpage
\section{Conclusion}\label{sec:con}
We combined different in-situ and migration models from the Genesis database and subjected the simulated planets to atmospheric mass-loss mechanisms including photo-evaporation and impact stripping. By comparing the model outcomes with the Kepler size distribution and to the observed sample of planets with precise masses and radii, we find that:
\begin{itemize}
\item Both in-situ and migration models are consistent with the radius valley and the mass-radius relations of the observed planets, but the migration models predict a significant number of water-rich planets within the orbital periods of 100 days while the in-situ models dictates that all of them are rocky.
\item Impact stripping alone can explain the size bimodality only with the migrating planets from the Genesis database, for a certain distribution of core mass and position. In such a case, impact stripping would result in a very high number of bare water planets which would be an indicator of this process.
\item Migration + photo-evaporation models, on the other hand, predict an intermediate fraction of bare planets to be water-rich ($\sim10$-30 \% around G-dwarfs and $\sim20$-35 \% around M-dwarfs) that are consistent with maximum possible fraction of water worlds from current observed sample of planets with precise mass and radius measurements ($24.2\pm6.1$ \% around G-dwarfs and $23.3\pm5.7$ \% around M-dwarfs).
\item However most bare water worlds are predicted from the migration models at orbital periods of  10-50 days where few RV follow-up efforts are concentrated. Similarly, the fraction of water worlds among the sub-Neptunes also  increases significantly outside of $\sim10$ days.
\item The code, Genesis Population Synthesis (\textit{GPS}), that can be used to develop population synthesis models based on the Genesis database of formation models and to reproduce the results of this paper can be found at: \url{https://github.com/arcunique/GPS}.
\end{itemize}

All the models have been generated with the help of the existing formation models of the Genesis database as our first step to developing the Genesis population synthesis models. Now that we have an end to end model of population synthesis, we plan to conduct more N-body simulations with a refined set of initial parameters based on the observed mass-radius-period distribution in our follow-up work. Based on our present study, we propose that follow-up radial velocity and spectroscopic surveys target planets at longer orbital periods to test the water world hypothesis, and provide a list of probable targets with high TSM and RV semi-amplitudes for follow-up.

We are thankful to the anonymous reviewer for having a critical reading of the manuscript and providing useful suggestions. A.C. acknowledges support from ANID -- Millennium Science Initiative -- ICN12\_009 -- Data Observatory Foundation. G.D.M. acknowledges support from FONDECYT project 11221206, from ANID --- Millennium Science Initiative --- ICN12\_009, and the ANID BASAL project FB210003. The results reported herein benefited from collaborations and/or information exchange within NASA’s Nexus for Exoplanet System Science (NExSS) research coordination network sponsored by NASA’s Science Mission Directorate and project “Alien Earths” funded under Agreement No. 80NSSC21K0593. A.C. further acknowledges his visit to the University of California, Santa Cruz and thanks Natalie Batalha, Artem Aguichine, and Anne Dattilo for all the valuable discussions and suggestions. This work also benefited from the 2023 Exoplanet Summer Program in the Other Worlds Laboratory (OWL) at the University of California, Santa Cruz, a program funded by the Heising-Simons Foundation and NASA.

\bibliography{ms}

\end{document}